\documentclass[aps,physrev,reprint,superscriptaddress,amsmath,amssymb]{revtex4-2}
\usepackage{physics}
\usepackage{dsfont}  
\usepackage{graphicx}
\usepackage[table]{xcolor}
\usepackage{ulem}
\usepackage{multirow}
\usepackage{placeins}
\usepackage{rotating}
\usepackage[
  colorlinks=true,
  linkcolor=blue,
  citecolor=blue,
  urlcolor=blue
]{hyperref}
\newcommand{\ts}{\textsuperscript} %
\newcommand{\bs}{\textsubscript} %
\newcommand{\Level}[3]{\ts{#1}#2\bs{#3}} %

\begin{document}

\title{Optical characterization of excited-state spin Hamiltonians and clock transitions at telecommunication wavelength in Tm$^{3+}$:YAlO$_{3}$}

\author{Akshay Babu Karyath}
\altaffiliation{These authors contributed equally to this work.}
\affiliation{Department of Applied Physics, University of Geneva, 1211 Geneva, Switzerland}

\author{Luozhen Li}
\altaffiliation{These authors contributed equally to this work.}
\affiliation{Department of Applied Physics, University of Geneva, 1211 Geneva, Switzerland}

\author{Julien Bertrand}
\altaffiliation{These authors contributed equally to this work.}
\affiliation{Department of Applied Physics, University of Geneva, 1211 Geneva, Switzerland}

\author{Alexandre Kings}
\affiliation{Department of Applied Physics, University of Geneva, 1211 Geneva, Switzerland}

\author{Tianyang Shen}
\affiliation{PSI Center for Photon Science, Paul Scherrer Institute, Villigen PSI, Switzerland}
\affiliation{Laboratory for Solid State Physics and Quantum Center, ETH Zurich, Zurich, Switzerland}

\author{Gabriel Aeppli}
\affiliation{PSI Center for Photon Science, Paul Scherrer Institute, Villigen PSI, Switzerland}
\affiliation{Laboratory for Solid State Physics and Quantum Center, ETH Zurich, Zurich, Switzerland}
\affiliation{Institute of Physics, EPF Lausanne, Lausanne, Switzerland}
\author{Simon Gerber}
\affiliation{PSI Center for Photon Science, Paul Scherrer Institute, Villigen PSI, Switzerland}

\author{Guy Matmon}
\affiliation{PSI Center for Photon Science, Paul Scherrer Institute, Villigen PSI, Switzerland}

\author{Wolfgang Tittel}
\email{Corresponding author: wolfgang.tittel@unige.ch}
\affiliation{Department of Applied Physics, University of Geneva, 1211 Geneva, Switzerland}

\date{\today}

\begin{abstract}
The zero-phonon line (ZPL) between the two excited-state manifolds \Level{3}{F}{4} and \Level{3}{H}{4} of the trivalent thulium ion in yttrium aluminum perovskite (Tm\ts{3+}:YAlO\bs{3}) lies within the telecom E-band at around 1451 nm and exhibits an optical coherence time of a few $\mu$s at cryogenic temperatures. To improve this time in view of potential applications, e.g. single-photon sources and optical quantum memory, we characterized the magnetic properties of the lowest Stark levels of these two manifolds. %
Varying the external magnetic field amplitude and orientation, we measured the enhanced nuclear Zeeman splitting by probing the field-dependent frequency detuning of side holes created by means of spectral hole burning, and we characterized the inhomogeneous line shift caused by the quadratic Zeeman interaction. Together, this allowed us to establish the spin Hamiltonians for the lowest Stark level of the \Level{3}{F}{4} and \Level{3}{H}{4} manifolds, and to predict optical clock transitions for particular orientations and magnitudes of the magnetic field.
\end{abstract}

\maketitle

\section{\label{sec:introduction}Introduction}
Trivalent rare-earth ions doped into inorganic crystals have become a promising platform for quantum communication technologies \cite{tittel2025quantum} due to their optical and spin coherence times at cryogenic temperatures that can reach milliseconds \cite{PhysRevLett.72.2179, PhysRevB.79.11510, PhysRevX.10.031060} and hours \cite{zhong2015optically, PRXQuantum.6.010302}, respectively. %
These  properties enable ensemble-based quantum memories, which have already allowed early demonstrations of quantum repeater infrastructure \cite{RevModPhys.83.33, PhysRevLett.113.053603, liu2021heralded, Lago_Rivera2021, Ortu2022, zhu2026metropolitan}. Beyond ensembles, individual rare-earth ions have emerged as single-photon sources \cite{dibos2018,zhong2018,yu2023frequency} and qubits for quantum information processing \cite{Raha2020, Kindem2020, Gritsch2025}.

Among the rare-earth ions, Tm$^{3+}$ is of significant %
interest. It is a non-Kramers ion with an even number of 4f electrons and only one stable isotope, \ts{169}Tm, with a nuclear spin of $I=1/2$. This results in a simple hyperfine structure. The optical ZPLs between the \Level{3}{H}{6} $\leftrightarrow$ \Level{3}{H}{4}, \Level{3}{F}{4} $\leftrightarrow$ \Level{3}{H}{4}, and \Level{3}{H}{6} $\leftrightarrow$ \Level{3}{F}{4} manifolds lie near 795 nm, 1450 nm, and 1780 nm, respectively, and are conveniently accessed using commercial semiconductor lasers. See Fig.~\ref{fig:energy_levels} for a level scheme.
The 795 nm \Level{3}{H}{6} $\leftrightarrow$ \Level{3}{H}{4} ground-state transition %
has received by far the most attention to date. It can exhibit very long optical coherence time, e.g. 1.1 ms in thulium-doped yttrium gallium garnet \cite{PhysRevB.90.214301}, and has been used for several demonstrations of quantum memory \cite{PhysRevLett.113.160501, PhysRevLett.127.220502}.

The \Level{3}{F}{4} $\leftrightarrow$ \Level{3}{H}{4} ZPL is very interesting due its wavelength of around 1451 nm where transmission loss in standard telecommunication fibers is small. As a first characterization of this transition, we have recently reported an optical coherence time of 4.75 $\mu$s in thulium-doped yttrium aluminum perovskite (Tm\ts{3+}:YAlO\bs{3}) \cite{li2026spectroscopy}. %
However, to enable the creation of long-lived optical quantum memory and coherent single-photon sources, it is important to further extend this time. %
To this end, one must suppress the effect of magnetic noise (from thulium-thulium interactions and from flipping spins in the host crystal) on the optical transition frequency. A well-established approach for spin transitions is to operate at zero-first-order-Zeeman (ZEFOZ) points: specific magnitudes and orientations of the applied magnetic field at which the transition frequency is first-order insensitive to magnetic-field fluctuations \cite{PhysRevLett.92.077601, zhong2015optically, PRXQuantum.6.010302}. Applying the analogous condition to the optical domain leads to an optical clock transition \cite{PhysRevA.85.032339, tongning2015, nicolas2023coherent, PhysRevB.104.134103}. To locate such %
transitions, we need full knowledge of the spin Hamiltonians of the two relevant energy levels. %

Spin Hamiltonians of rare-earth ions have been characterized previously using a variety of techniques, such as spectral hole burning (SHB)\cite{macfarlane1981measurement, PhysRevB.73.085112}, electron paramagnetic resonance (EPR)\cite{PhysRevB.74.214409}, and optical spectroscopy combined with radio-frequency or microwave resonance measurements \cite{PhysRevB.74.195101, PhysRevB.84.104417, PhysRevB.85.014429, PhysRevB.98.195110, PhysRevB.104.134103}. Here we use spectral holes burned in the inhomogeneously broadenend 1451 nm transition to characterize the spin Hamiltonians of the lowest Stark levels of the \Level{3}{F}{4} and \Level{3}{H}{4} manifolds in Tm\ts{3+}:YAlO\bs{3}, and we also predict optical clock transitions between these two excited states.

The paper is organized as follows. Section~\ref{sec:spin_hamiltonian} gives an overview of the spin Hamiltonian for a non-Kramers ion in a crystal and introduces the parameters to be characterized. Section~\ref{sec:crystal_symmetry} describes the symmetry of the Tm\ts{3+}:YAlO\bs{3} crystal and the sites occupied by the thulium ions. Section~\ref{sec:setup} details the experimental setup. Section~\ref{sec:results} presents measurement results and the analysis of the spectral hole burning and inhomogeneous line shift data. Section~\ref{sec:optical_clock_transitions} discusses our predictions for optical clock transitions based on the characterized spin Hamiltonians. Finally, section~\ref{sec:conclusion} summarizes the paper. 

\section{\label{sec:spin_hamiltonian}Spin Hamiltonian}
The energy levels of a non-Kramers ion in a crystal-field environment are usually governed by a Hamiltonian of the following form \cite{abragam2012electron}:

\begin{equation}
    H = (H_{\text{FI}} + H_{\text{CF}}) + H_{\text{Q}} + H_{\text{HF}} + H_{\text{eZ}} + H_{\text{nZ}}.
    \label{eq:spin_Hamiltonian}
\end{equation}
\noindent
The first term ($H_\text{FI}$) denotes the free-ion Hamiltonian, comprising the Coulomb and spin-orbit interactions that split the 4f\ts{12} configuration into manifolds denoted \Level{2S+1}{L}{J}, where S is the total spin angular momentum, L is the total orbital angular momentum, and J is the total angular momentum.   %
 When only considering the free-ion Hamiltonian, each manifold remains degenerate. The second term ($H_\text{CF}$) describes the crystal-field Hamiltonian, which is determined by the host crystal. If a non-Kramers ion such as Tm\ts{3+} occupies a site with less than threefold rotational symmetry \cite{PhysRevB.73.085112, abragam2012electron}, the crystal field lifts the degeneracy of each manifold, splitting them into non-degenerate singlets that are known as Stark levels. The first two terms in Eq. (\ref{eq:spin_Hamiltonian}) determine the energy of the Stark levels and have been characterized in previous works \cite{antonov1973, Ohare1976, korner2018spectroscopic}. 

The third term describes the  quadrupole Hamiltonian ($H_\text{Q}$).  But since the electric quadrupole interaction requires a nuclear spin $I\geq1$ \cite{abragam1961principles}, it vanishes for thulium where $I=1/2$. 

The remaining three terms -- hyperfine Hamiltonian ($H_\text{HF}$), electronic Zeeman Hamiltonian ($H_\text{eZ}$), and nuclear Zeeman Hamiltonian ($H_\text{nZ}$) -- are treated as perturbations of the Stark levels. The hyperfine Hamiltonian describes the magnetic coupling between the electronic and nuclear spins of thulium. The electronic Zeeman Hamiltonian describes the coupling between the electronic spin of thulium and the external magnetic field. Finally, the nuclear Zeeman Hamiltonian describes the coupling between the nuclear spin of thulium and the external magnetic field. Combining these three terms, we define the perturbation Hamiltonian $H_{\text{pert}}$ as follows:

\begin{equation}
\begin{aligned}
    H_{\text{pert}} &= H_{\text{HF}} + H_{\text{eZ}} + H_{\text{nZ}} \\
    &= A \mathbf{I} \cdot \mathbf{J} + g_{J} \mu_{B}\mathbf{J} \cdot \mathbf{B} - g_{n} \mu_{N} \mathbf{I} \cdot \mathbf{B}.
\end{aligned}
\label{eq:pert_Hamiltonian}
\end{equation}
Here, $\mathbf{I}$ is the nuclear spin operator, $\mathbf{J}$ is the total electronic angular momentum operator, $A$ is the magnetic hyperfine constant, $g_{J}$ is the Landé g-factor, $g_{n}$ is the nuclear g-factor, $\mu_{B}$ is the Bohr magneton, $\mu_{N}$ is the nuclear magneton, and $\mathbf{B}$ is the applied magnetic field.

\begin{figure*}[t]
    \centering
    \includegraphics[width=1.4 \columnwidth]{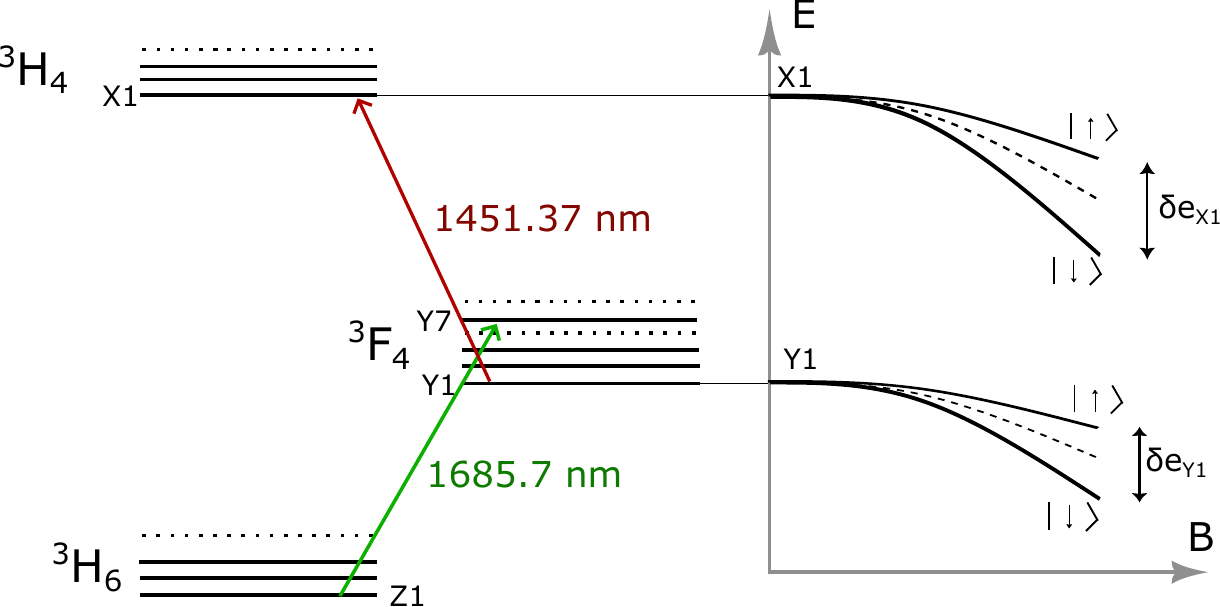}
    \caption{Energy level diagram of Tm\ts{3+}:YAlO\bs{3} where \Level{3}{H}{6} is the ground-state manifold ; \Level{3}{F}{4} and \Level{3}{H}{4} are excited-state manifolds. At right, a schematic of the evolution of the lowest Stark levels of the \Level{3}{F}{4}  and \Level{3}{H}{4} manifolds (Y1 and X1, respectively) with applied magnetic field.  Each level shifts quadratically (dashed line, due to the quadratic Zeeman shift) and splits linearly into two nuclear-spin sublevels (solid line, caused by the enhanced nuclear Zeeman effect), with total splittings $\delta e_{\mathrm{X1}}$ and $\delta e_{\mathrm{Y1}}$, respectively. $\ket{\uparrow}$ and $\ket{\downarrow }$ denotes the nuclear spin states. Energy axes not to scale.}
    \label{fig:energy_levels}
\end{figure*}

First, we consider the first-order contribution of $H_{\text{pert}}$. Because thulium is a non-Kramers ion and each Stark level in a low symmetry host is a non-degenerate singlet, it therefore carries no electronic magnetic moment ($\bra{i}\mathbf{J}\ket{i}=0$ for any Stark level $\ket{i}$) \cite{abragam2012electron}. The hyperfine and electronic Zeeman terms are both proportional to $\mathbf{J}$ and therefore vanish at first order. Then, only the nuclear Zeeman Hamiltonian contributes at first order, yielding $H_{\text{eff}}^{(1)}= - g_{n} \mu_{N} \mathbf{I} \cdot \mathbf{B}$.

We furthermore consider the second-order contributions $H_{\text{eff}}^{(2)}$ from the hyperfine and electronic Zeeman terms. The full effective Hamiltonian can then be written as
\begin{equation}
\begin{aligned}
H_{\mathrm{eff}}
&= H_{\mathrm{eff}}^{(1)} + H_{\mathrm{eff}}^{(2)} \\
&= \mathbf{B}^{\mathsf T}
\left(
-g_n\mu_N\mathds{1}
-2g_J\mu_BA\mathbf{\Lambda}
\right)\mathbf{I}
-g_J^2\mu_B^2
\mathbf{B}^{\mathsf T}\mathbf{\Lambda}\mathbf{B}.
\end{aligned}
\label{eq:perturbation_theory_Hamiltonian}
\end{equation}
where $\mathbf{\Lambda}$ is a tensor, defined as

\begin{equation}
    \Lambda_{\alpha\beta} = \sum_{n \neq 1} \frac{\langle 1 | J_\alpha | n \rangle \langle n | J_\beta | 1 \rangle}{E_n - E_1}, \qquad \alpha, \beta \in \{x, y, z\}
    \label{eq:lambda_tensor}
\end{equation}
\noindent
and ($x,y,z$) denote the principal axes of the tensor. The $\mathbf{\Lambda}$ tensor describes how the total electronic angular momentum operator $\mathbf{J}$, which enters both the electronic Zeeman and hyperfine interactions, mixes the lowest Stark level ($\ket{1}$) with the other Stark levels. $E_{n}-E_{1}$ denotes the energy difference between the Stark levels $\ket{n}$ and $\ket{1}$. See Appendix~\ref{app:perturbation theory} for detailed derivations of Eq.~\eqref{eq:perturbation_theory_Hamiltonian}.

The first term in Eq.~\eqref{eq:perturbation_theory_Hamiltonian} describes the enhanced nuclear Zeeman effect, comprising a part due to the first-order nuclear Zeeman interaction ($-g_{n}\mu_{N}\mathds{1}$) and a second-order contribution from the combination of electronic Zeeman and hyperfine interactions ($-2g_{J}\mu_{B}A\mathbf{\Lambda}$). The second term in Eq.~\eqref{eq:perturbation_theory_Hamiltonian} describes the quadratic Zeeman effect, which results from the second-order contribution of the electronic Zeeman interaction ($-g_J^2\mu_B^2\mathbf{B}^{\mathsf T}\mathbf{\Lambda}\mathbf{B}$). 

As illustrated in Fig.~\ref{fig:energy_levels}, each Stark level shifts under an applied magnetic field quadratically through the quadratic Zeeman effect and splits linearly into its two nuclear-spin sublevels through the enhanced nuclear Zeeman effect.

\begin{figure}[t]
    \centering
    \includegraphics[width=0.9\columnwidth]{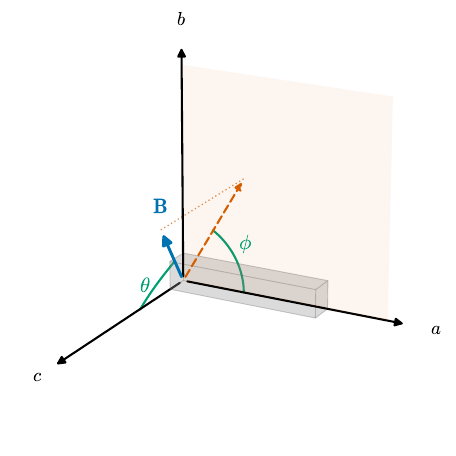}
    \caption{Definition of the magnetic-field coordinate system relative to the crystal axes (a, b, c) of Tm\ts{3+}:YAlO\bs{3}. The magnetic field $\mathbf{B}$ is parametrized by the polar angle $\theta$, measured with respect to the $c$ axis and the azimuthal angle $\phi$, measured in the $ab$ plane with respect to the $a$ axis. The gray box indicates the cut of the Tm\ts{3+}:YAlO\bs{3} crystal.}
    \label{fig:coordinate_system}
\end{figure}

\section{\label{sec:crystal_symmetry}Crystal and site symmetry }

The crystal structure of YAlO\bs{3} belongs to the orthorhombic space group $Pbnm$, with lattice constant $a=5.179~\mathrm{\AA}$, $b=5.329~\mathrm{\AA}$, $c=7.370~\mathrm{\AA}$, where $a$, $b$ and $c$ are the crystal axes. 
 Tm\ts{3+} ions substitute Y\ts{3+} ions located at 4c Wyckoff positions \cite{ju2016structural}. These 4 sites (c1, c2, c3, c4) can be grouped in magnetically equivalent pairs, because site c3 (resp. c4) can be deduced from site c1 (resp. c2) by the inversion symmetry of the space group \cite{antonov1973}. In the following, we call the pair (c1, c3) site 1, and the pair (c2, c4) site 2. Site 2 can be deduced from site 1 by a gliding plane mirror symmetry along the $b$ axis (or $a$ axis). Thus, sites 1 and 2 become magnetically equivalent for magnetic fields located in the principal planes $ac$ and $bc$. This was exploited to describe the reference frame of the magnetic field in the reference frame of the crystal; the method is detailed in the Appendix~\ref{app:calib}.

Each individual site exhibits $C_{s}$ symmetry, i.e. a mirror plane orthogonal to the $c$ axis. We denote $\mathbf{\Lambda}_{l}^{i}$  the $\mathbf{\Lambda}$ tensor of site $i$ ($i$=1,2) and level $l$ ($l$=Z1, Y1, X1). By application of Neumann's principle (see Appendix \ref{app:sym details} for more details), we find:\\
\begin{align}
    \mathbf{\Lambda}_l^{i} &=
    \begin{pmatrix}
        \Lambda_{l_{cc}} & 0 & 0\\
        0 & \Lambda_{l_{aa}}^{i}& \Lambda_{l_{ab}}^{i}\\
        0 & \Lambda_{l_{ba}}^{i} & \Lambda_{l_{bb}}^{i}
    \end{pmatrix}_{(c,a,b)} \nonumber\\
    &=
    \begin{pmatrix}
        \Lambda_{l_{xx}} & 0 & 0\\
        0 & \Lambda_{l_{yy}}& 0\\
        0 & 0 & \Lambda_{l_{zz}}
    \end{pmatrix}_{(x,y,z)_l^{i}}
    \label{eq:one}
\end{align}
\noindent
where ($x,y,z$) refers to the principal axes of the $\mathbf{\Lambda}$ tensor. Note that this diagonalization basis is different for each energy level $l$ and for each site $i$ and that it also differs from the crystal axes ($a$, $b$, $c$). However, the $C_s$ symmetry forces the $x$ axis of sites 1 and 2 to coincide with the $c$ axis of the crystal. Hence,  the axes $y$ and $z$ are contained within the $ab$ plane. 

We furthermore define $\alpha_l$ as the angle between the $a$ axis and the $y$ axis of site 1. Then, the glide plane mirror along the $a$ axis imposes the angle between the $a$ axis and the $y$ axis of site 2 to be $-\alpha_l$ \cite{asatryan2002epr}; for more details, see Appendix \ref{app:sym details}.

\section{\label{sec:setup}Experimental Setup}

Measurements were carried out using a 0.1\% Tm\ts{3+}-doped YAlO\bs{3} crystal (dimensions 12.0 $\times$ 2.2 $\times$ 2.0 $\mathrm{mm}^3$ along the $a$, $b$, and $c$ axes, respectively; Scientific Materials Corp.). The crystal was thermally glued to a Cu stage using silver paste, and further secured by a Cu lid with spring-tensioned screws. The crystal, along with this stage was cooled down to around 800 mK using an Oxford Instruments cryogen-free Triton \ts{3}He-\ts{4}He dilution refrigerator. A superconducting split-coil vector magnet allows applying a homogeneous magnetic field of up to 2 T along arbitrary directions with respect to the crystal axes. The flux density homogeneity of all coils is better than 1.8 \% over the crystal volume. As shown in Fig.~\ref{fig:coordinate_system}, we define a spherical coordinate system for the magnetic field%
: the polar angle $\theta$ is the angle between the field vector and the crystal's $c$ axis, and the azimuthal angle $\phi$ is the angle between the projection of the field onto the $ab$ plane and the crystal's $a$ axis.

A tunable continuous-wave (CW) laser (Toptica DL Pro) emitting at around 1685 nm was used for optical pumping of Tm\ts{3+}ions from the \Level{3}{H}{6} level into the \Level{3}{F}{4} level, and another CW laser (Santec TSL-550) probed the excited-state transition at around 1450 nm. Fiber based acousto-optic modulators (AOMs) were used for creating short laser pulses, and a phase modulator (PM) was used to scan the frequency of the probe pulse via a serrodyne signal \cite{johnson2010broadband}. Separate fiber polarization controllers were used for the two lasers since pumping and probing are polarization dependent. After the polarization controllers, the lasers were combined using a fiber beam splitter whose output was connected to a free-space setup to focus the beam into the cryostat. This setup was composed of 
two achromatic doublets with focal lengths of 60 mm and 200 mm (Thorlabs AC254-060-C-ML and AC254-200-C-ML, respectively), which were used as a telescope to expand the beam, and a third achromatic doublet of focal length 500 mm (AC254-500-C-ML) that focused the light into the crystal where it propagated along the $a$ axis. A periscope allowed translating the beam horizontally and vertically, and the focusing lens was mounted on a translation stage to tune the beam-waist position relative to the crystal. Note that we used achromatic doublets to minimize the focus shift between pump and probe beams, and that all lenses were located  outside the cryostat. Behind the cryostat, the light was refocused using a 500 mm achromatic lens, and coupled into a single mode telecom fiber before being detected by a 10 MHz-bandwidth InGaAs photoreceiver (New Focus 2053). A schematic of the set-up is given in Appendix \ref{app:set up}, see also \cite{li2026spectroscopy} for details.

\begin{figure}[t]
    \centering
    \includegraphics[width=0.9\columnwidth]{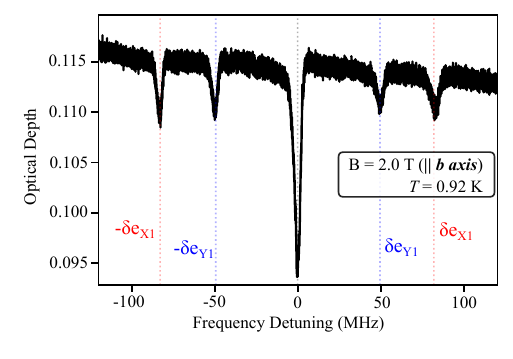}
    \caption{Spectral holes of the 1451 nm Y1-X1 transition at $T$=0.92~K with an applied magnetic field of 2~T oriented along the b-axis of the crystal. The figure shows a central hole at zero detuning together with four side holes located symmetrically about it, which arise from the inhomogeneous broadening of the transition along with atomic level splittings. When $\mathbf{B}\parallel b$, the detunings of the inner pair (blue dotted lines) and the outer pair (red dotted lines) correspond to the enhanced nuclear Zeeman splittings $\delta e_{\mathrm{Y1}}$ and $\delta e_{\mathrm{X1}}$ of the Y1 and X1 levels, respectively.}
    \label{fig:sideholes}
\end{figure}

\section{\label{sec:results}Measurements and results}

\subsection{\label{sec:SHB}Spectral hole burning}
To assess magnetic-field-dependent level shifts and level splitting, we measured the wavelength of the center of the inhomogeneously broadened Y1 $\leftrightarrow$X1 absorption line and the detuning of side-holes created in the absorption profile by means of spectral hole burning \cite{li2026spectroscopy}, respectively. In either case, we first use the 1685 nm laser to pump atomic population from the lowest Stark level of the \Level{3}{H}{6} manifold (Z1) to the 7th Stark level of the \Level{3}{F}{4} manifold (Y7), from where it rapidly decays into Y1 since the non-radiative decay rate depends exponentially on the energy-gap between stark levels \cite{riseberg1968multiphonon}. Next, we use the 1451 nm laser to either scan across the Y1$\leftrightarrow$X1 line and establish its center, or to excite some of the ions over a narrow spectral interval into the lowest level (X1) of the \Level{3}{H}{4} manifold. This leads to a modification of the absorption profile, a so-called spectral hole. After a wait time that is shorter than the 0.85 ms lifetime of this level and during which no laser interacts with the ions, we switch the 1451 nm laser back on and scan it over a spectral interval of up to a few GHz, depending on the measurement, that is centered on the hole. We monitor the wavelength-dependent transmission as a function of detuning, i.e. the frequency difference of the laser during scanning with respect to the frequency used for hole burning. 

 For a general orientation of the applied magnetic field, each of the two magnetically inequivalent thulium sites gives rise to a main hole at zero detuning and four side holes. The two sites become magnetically equivalent whenever the field lies in the $ac$ or $bc$ plane, in which case their side holes coincide and only four such holes are resolved. See Fig. \ref{fig:sideholes} for an example with $\mathbf{B}\parallel b$. As detailed in the Supplemental Material of Ref.~\cite{li2026spectroscopy}, time-resolved spectral hole burning allows one to match each side hole to its corresponding level, indicating the splitting between the spin states of the X1 or the Y1 level (respectively the outer and inner pairs in Fig. \ref{fig:sideholes})).

For the X1 and Y1 levels, the unknowns of the spin-Hamiltonians are the $\mathbf{\Lambda}_l$ tensors and the hyperfine constants $A_l$. The two measurements described above yield information about level shifts and level splitting, respectively, and are complementary. For each level, the enhanced nuclear Zeeman splitting depends on the products $A_l \Lambda_{l,\sigma\sigma}$ and the orientation, $\alpha$, of the principal axes of $\mathbf{\Lambda}_l$. However, the quadratic Zeeman shift only depends on $\Lambda_{l,\sigma\sigma}$ and $\alpha$. Therefore, combining all measurements yields both the hyperfine constants $A_l$ and the principal values of $\mathbf{\Lambda}_l$.
 
\begin{figure*}[t]
  \centering
  \newcommand{\panelw}{0.315\textwidth}

  \makebox[\panelw]{$\mathbf{B}\in ab$ plane ($\theta=90^{\circ}$, $\phi$ scan)}\hfill
  \makebox[\panelw]{$\mathbf{B}\in bc$ plane ($\theta$ scan, $\phi=90^{\circ}$)}\hfill
  \makebox[\panelw]{$\mathbf{B}\in ac$ plane ($\theta$ scan, $\phi=0^{\circ}$)}\\[1pt]

  \begin{minipage}[t]{\panelw}\centering
    \includegraphics[width=0.9\linewidth]{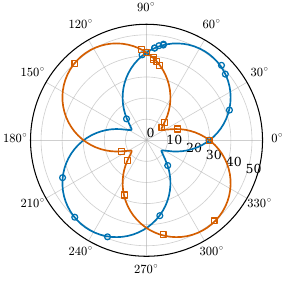}
    \\[-3pt]{\small (a) X1}
  \end{minipage}\hfill
  \begin{minipage}[t]{\panelw}\centering
    \includegraphics[width=0.9\linewidth]{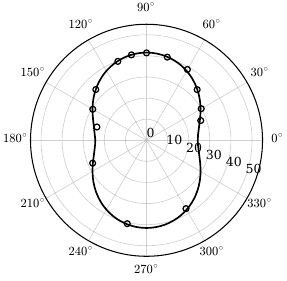}
    \\[-3pt]{\small (b) X1}
  \end{minipage}\hfill
  \begin{minipage}[t]{\panelw}\centering
    \includegraphics[width=0.9\linewidth]{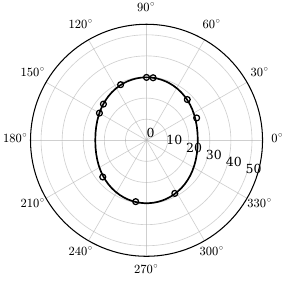}
    \\[-3pt]{\small (c) X1}
  \end{minipage}\\[3pt]

  \begin{minipage}[t]{\panelw}\centering
    \includegraphics[width=0.9\linewidth]{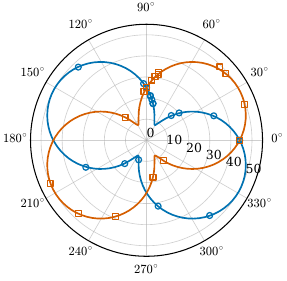}
    \\[-3pt]{\small (d) Y1}
  \end{minipage}\hfill
  \begin{minipage}[t]{\panelw}\centering
    \includegraphics[width=0.9\linewidth]{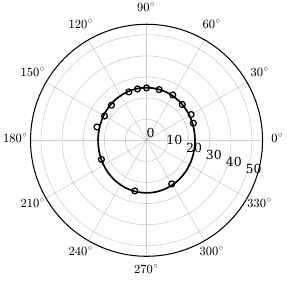}
    \\[-3pt]{\small (e) Y1}
  \end{minipage}\hfill
  \begin{minipage}[t]{\panelw}\centering
    \includegraphics[width=0.9\linewidth]{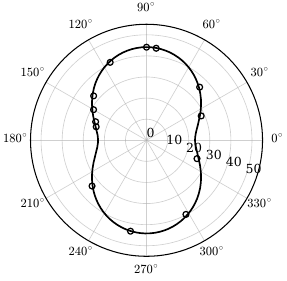}
    \\[-3pt]{\small (f) Y1}
  \end{minipage}

  \caption{Angular dependence of the enhanced nuclear Zeeman splitting rate
 for the X1 level of the $^{3}\mathrm{H}_{4}$ manifold and the Y1 level of the
    $^{3}\mathrm{F}_{4}$ manifold of Tm$^{3+}$:YAlO$_3$. Top row
    (a)-(c): X1; bottom row (d)-(f):
    Y1. For $\mathbf{B}$ within the $ab$ plane (panels (a),(d)) site~1 and site~2 are magnetically inequivalent, resulting in distinct
    splittings (blue circles, site~1; orange squares, site~2), whereas for $\mathbf{B}$ within the $bc$ and $ac$ planes (panels (b),(e) and (c),(f), respectively), the sites are magnetically
    equivalent and yield the same splittings. Markers depict measured
    data; solid lines are fits using Eq.~\eqref{eq:splitting_value}. Radial axis: splitting rate in MHz/T. The procedure for assigning measurements to site 1 and site 2 is discussed in Appendix~\ref{app:site_convention}.}
  \label{fig:enhanced_nuclear_Zeeman}
\end{figure*}

\subsection{\label{sec:Zeeman splitting}Enhanced nuclear Zeeman splitting}

As discussed in Sec.~\ref{sec:spin_hamiltonian}, the enhanced nuclear Zeeman effect  splits each Stark level linearly into two nuclear-spin sublevels. 
For convenience, based on Eq.~\eqref{eq:perturbation_theory_Hamiltonian}, we define the enhanced nuclear Zeeman tensor $\mathbf{\Gamma}=-g_{n}\mu_{N}\mathds{1}-2g_{J}\mu_{B}A\mathbf{\Lambda}$. Since the two magnetically inequivalent sites are related by a glide plane mirror, their $\mathbf{\Gamma}$ tensors have the same eigenvalues and the principal axes are related by ($\alpha\to-\alpha$) (see Sec.~\ref{sec:crystal_symmetry} and Appendix \ref{app:sym details}). We therefore consider only site 1 in the following equations.

In the $(x,y,z)_{l}$ basis, the enhanced nuclear Zeeman tensor $\mathbf{\Gamma}_{l}$ is  diagonal with elements
\begin{equation}
    \gamma_{l,\sigma\sigma}=-g_{n}\,\mu_{N}-2g_{J,l}\,\mu_{B}\,A_{l}\,\Lambda_{l,\sigma\sigma},
    \label{eq:diagonal_terms_enhanced_nuclear_zeeman}
\end{equation}
where $\sigma \in \{x, y, z\}$ labels the principal axes of the $\Lambda$ tensor and $l \in \{\mathrm{X1}, \mathrm{Y1}\}$ the two Stark levels. Furthermore, the applied magnetic field has components
\begin{equation}
    \mathbf{B}=B\begin{pmatrix}
        \cos(\theta) \\
        \sin(\theta)\cos(\phi-\alpha_{l})\\
        \sin(\theta)\sin(\phi-\alpha_{l})
    \end{pmatrix},
    \label{eq:B_field_xyz}
\end{equation}
where $\theta$ and $\phi$ are the crystal-frame angles defined in Fig.~\ref{fig:coordinate_system}. Substituting Eq.~\eqref{eq:diagonal_terms_enhanced_nuclear_zeeman} and \eqref{eq:B_field_xyz} into Eq.~\eqref{eq:perturbation_theory_Hamiltonian}, the enhanced nuclear Zeeman splitting rate for level $l$ under external magnetic field $\mathbf{B}$ can be written as 
\begin{multline}
    \delta e_{l}/B = [ \gamma_{l,xx}^2 \cos^2\theta
    + \gamma_{l,yy}^2 \sin^2\theta \cos^2(\phi-\alpha_l) \\
    + \gamma_{l,zz}^2 \sin^2\theta \sin^2(\phi-\alpha_l) ] ^{1/2}.
    \label{eq:splitting_value}
\end{multline}

We measured the enhanced nuclear Zeeman splittings $\delta e_{l}$ for $l\in{\text{X1},\text{Y1}}$ for varying orientations and amplitudes of the magnetic field. %
The splitting rates $\delta e_{l}/B$, depicted in Fig.~\ref{fig:enhanced_nuclear_Zeeman}, are fitted using Eq.~\eqref{eq:splitting_value}, where $\gamma_{l,xx}$, $\gamma_{l,yy}$, $\gamma_{l,zz}$ and $\alpha_l$ are treated as free parameters. For each level $l$, we simultaneously fit the splitting rates measured when $\mathbf{B}$ lies within the $ab$ plane (both sites included), and the $bc$ and $ac$ planes (where both sites are magnetically equivalent) using Eq.~\eqref{eq:splitting_value} and a single set of parameters. Due to the mirror symmetry, site 2 is constrained to have the same values for $\gamma_{l,xx}$, $\gamma_{l,yy}$ and $\gamma_{l,zz}$, and the opposite sign for its principal-axis angle $-\alpha_l$; this symmetry is directly imposed in the fitting algorithm. Since the parameter sets ($\gamma_{l,xx}$, $\gamma_{l,yy}$, $\gamma_{l,zz}$, $\alpha_l$) and ($\gamma_{l,xx}$, $\gamma_{l,zz}$, $\gamma_{l,yy}$, $\alpha_l \pm 90^{\circ}$) describe the same angular dependence of the enhanced nuclear Zeeman splitting, we adopt the convention $\gamma_{l,yy}<\gamma_{l,zz}$ to remove the ambiguity.

The fitted parameters for X1 and Y1 are listed in  Table~\ref{tab:enhanced_nuclear_Zeeman_global_fit}. The quality of the fits can be assessed by comparing the residual standard error (RSE) with the scale of the measured values, showing excellent agreement. See the Appendix \ref{app:fit} for more information.

\begin{table}[t]
  \centering
  \caption{Principal values of the enhanced nuclear Zeeman tensor $\mathbf{\Gamma}_l$ and the in-plane principal-axis angle $\alpha$ for the lowest Stark levels of $^{3}\mathrm{H}_{4}$ (X1) and $^{3}\mathrm{F}_{4}$ (Y1) of Tm$^{3+}$:YAlO$_3$.}
  \label{tab:enhanced_nuclear_Zeeman_global_fit}
  \begin{ruledtabular}
  \begin{tabular}{lcc}
     & X1 ($^{3}\mathrm{H}_{4}$) & Y1 ($^{3}\mathrm{F}_{4}$) \\
    \colrule
    $\gamma_{xx}/h$ (MHz/T) & 24.30(16) & 22.98(16) \\
    $\gamma_{yy}/h$ (MHz/T) &  8.66(20) &  8.23(25) \\
    $\gamma_{zz}/h$ (MHz/T) & 50.44(8) & 50.05(11) \\
    $\alpha$ (deg), site 1  & -35.08(10) & 61.52(11) \\
    $\alpha$ (deg), site 2  & 35.08(10)  & -61.52(11) \\
  \end{tabular}
  \end{ruledtabular}
\end{table}

\subsection{\label{sec:quadratic_shift}Quadratic Zeeman effect}

\begin{figure}[t]
    \centering
    \includegraphics[width=0.9\columnwidth]{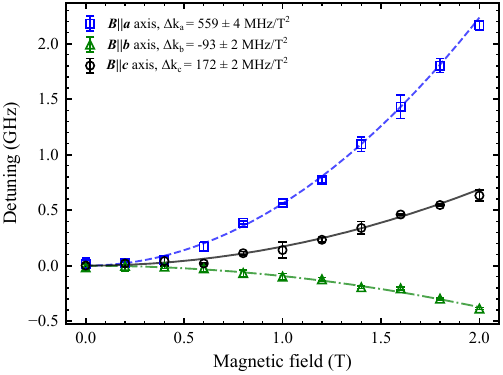}
    \caption{Detuning of the center of the inhomogeneous broadened line of the excited-state transition (Y1 - X1) as a function of magnetic field amplitude, for $\mathbf{B}$ oriented along the three crystal axes of Tm\ts{3+}:YAlO\bs{3}. Each set of data is fitted with a quadratic function $\Delta \nu=k_i \cdot B^{2}$ (i=a,b,c), after subtracting the zero-field ($B = 0\,\mathrm{T}$) baseline detuning. The fitted quadratic shift rates $\Delta k_i$ are indicated, including uncertainties. The error bars of the individual data points characterize the uncertainty with which the center of the inhomogeneous absorption line could be established.}
    \label{fig:quad_zeeman_shift}
\end{figure}

\begin{figure}[t]
    \centering
    \small{(a) $\mathbf{B}\in ab$ plane ($\theta=90^{\circ}$, $\phi$ scan)}
    \includegraphics[width=0.6\columnwidth]{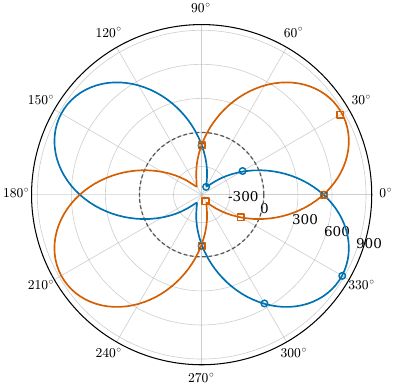}
    \vspace{0.5cm}
    
    \small{(b)$\mathbf{B}\in ac$ plane ($\theta$ scan, $\phi=0^{\circ}$)}
    \includegraphics[width=0.6\columnwidth]{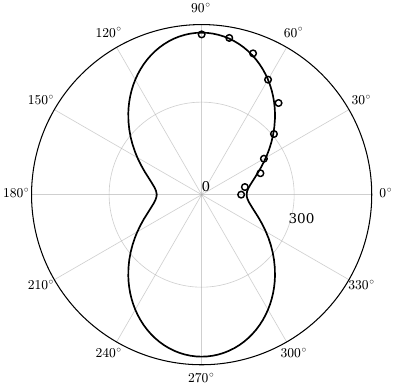}
    
    \caption{Angular dependence of the quadratic Zeeman shift of the Y1 to X1 transition frequency. (a) Scan of the magnetic field in the $ab$ plane (blue circles, site~1; orange squares, site~2). (b) Scan of the magnetic field in the $ac$ plane (where the two sites are magnetically equivalent). Radial axis: $k_{\text{X1}}-k_{\text{Y1}}$ in MHz/T\ts{2}. Solid lines are the fitted curves. }
    \label{fig:orient shift}
\end{figure}

According to Eq.~\eqref{eq:perturbation_theory_Hamiltonian}, each Stark level shifts quadratically with the magnetic-field magnitude at a rate $k_l$ set by the projection of the $\mathbf{\Lambda_l}$ tensor onto the field direction. These rates rate can be expressed using
\begin{multline}
    k_l = \frac{g_{J,l}^2 \mu_B^2}{h} [\Lambda_{l_{xx}} \cos^2{(\theta)}+\Lambda_{l_{yy}} \sin^2{(\theta}) \cos^2{(\phi-\alpha_l)}\\+\Lambda_{l_{zz}} \sin^2{(\theta)}\sin^2{(\phi-\alpha_l)}].
    \label{eq:shift value}
\end{multline}
Note that both the Y1 level and the X1 level shift under an applied magnetic field; the shift of the Y1-X1 transition frequency therefore reflects the difference of their quadratic shift rates, $k_{\text{X1}}-k_{\text{Y1}}\equiv \Delta k$.

As shown in Fig.~\ref{fig:quad_zeeman_shift}, we measured the detuning of the center of the Y1-X1 absorption line as the magnetic-field magnitude was varied from 0 to 2 T, with the magnetic field applied along one of the three crystal axes ($a$, $b$, $c$). The frequency at B=0 T was taken as the reference point (zero detuning), and all measurements were done using a wavemeter (Bristol Instruments Model 671). The data are well fitted by a quadratic function, with fitted shift rates $\Delta k_a =559\pm4$, $\Delta k_b =-93\pm2$ and $\Delta k_c =172\pm2~\mathrm{MHz/T^2}$ for $\mathbf{B}\parallel$ $a$, $b$, $c$, respectively.

We furthermore measured the quadratic shift of the Y1-X1 transition frequency %
at B = 2~T for varying orientations of B, see Fig. \ref{fig:orient shift}. These results, along with the measured eigenvalues of the $\mathbf{\Gamma}$ tensor (see Table \ref{tab:enhanced_nuclear_Zeeman_global_fit}) allow us to establish the magnetic hyperfine constants $A_l$ for the \Level{3}{H}{4} and \Level{3}{F}{4} manifolds. Towards this end, we use Eq.~\eqref{eq:diagonal_terms_enhanced_nuclear_zeeman} to find expressions for $\Lambda_{l_{,\sigma\sigma}}$. In turn, this allows us to express $\Delta k$ using Eq.~\eqref{eq:shift value} in terms of $\gamma_{lxx}/h$, $\gamma_{lyy}/h$, $\gamma_{lzz}/h$ and $A_l$, where $h$ is the Planck constant. Treating $A_{\mathrm{3H4}}$ and $A_{\mathrm{3F4}}$ as free fit parameters, we fit $\Delta k$ simultaneously to the angular dependence of the quadratic Zeeman shift rates measured in the $ab$ plane, for both sites, and in the $ac$ plane, as shown in Figs.~\ref{fig:orient shift} (a) and (b).

The fit results in $A_{\mathrm{3H4}}/h = -465.54 \pm 4.43~\mathrm{MHz}$ and $A_{\mathrm{3F4}}/h = -430.05 \pm 2.33~\mathrm{MHz}$. Substituting the fitted magnetic hyperfine constant $A_l$ and the principal values of the enhanced nuclear Zeeman tensor $\mathbf{\Gamma_l}$ into Eq.~\eqref{eq:diagonal_terms_enhanced_nuclear_zeeman}, we can calculate the $\Lambda$ tensor for each X1 and Y1 level.

This completes the characterization of the spin Hamiltonian of the X1 and Y1 levels, listed in Table~\ref{tab:spin_hamiltonian_params_global_fit}, where the corresponding values for the Z1 level of the ground \Level{3}{H}{6} manifold are included for completeness using existing literature values \cite{PhysRevB.104.134103, PhysRevB.71.174409, asatryan2013wide}. The Landé g-factor $g_J$ of these three levels are obtained from the calculations in the Russel-Saunders coupling approximation \cite{abragam2012electron}.

We emphasize that the magnetic hyperfine constants $A_l$ of the \Level{3}{H}{4} and \Level{3}{F}{4} levels are supposed to be independent of the crystal host. These constants were calculated for both levels in a previous work \cite{PhysRevB.71.174409} and have been used to characterize spin-Hamiltonians in Tm\ts{3+}:Y$_{3}$Ga$_{5}$O$_{12}$ and Tm\ts{3+}:Y$_{3}$Al$_{5}$O$_{12}$ \cite{PhysRevB.71.174409, PhysRevB.104.134103}. Interestingly, our measured value of -465.5 MHz for the \Level{3}{H}{4} level differs from the calculated value of -678.3 MHz. Note that the sites occupied by Tm\ts{3+} ions in YAlO\bs{3} exhibit a very low point symmetry ($C_s$), in which the assumption of a radial 4f wavefunction that underpins the calculation may not hold \cite{freeman1962theoretical,owen1966covalent}.

\begin{table}[t]
 \centering
 \caption{Parameters of Tm\ts{3+}:YAlO\bs{3} spin Hamiltonian.}
 \label{tab:spin_hamiltonian_params_global_fit}
 \begin{ruledtabular}
 \begin{tabular}{lccc}
    & {X1 ($^{3}\mathrm{H}_{4}$)} & {Y1 ($^{3}\mathrm{F}_{4}$)} & {Z1 ($^{3}\mathrm{H}_{6}$)} \\
   \colrule
   $g_J$
     & 0.80 & 1.25 & 1.16 \\
   $A/h$ (MHz)
     & -465.5(44) & -430.1(23) & -470.3~\cite{PhysRevB.71.174409} \\
   $\Lambda_{xx}$ ($10^{21}\,\mathrm{J}^{-1}$)
     & 3.01(3)& 1.95(1)& {${\approx}\,0$}~\cite{asatryan2013wide} \\
   $\Lambda_{yy}$ ($10^{21}\,\mathrm{J}^{-1}$)
     & 0.743(7)& 0.472(3)& {${\approx}\,0$}~\cite{asatryan2013wide} \\
   $\Lambda_{zz}$ ($10^{21}\,\mathrm{J}^{-1}$)
     & 6.79(7)& 4.67(3)& 374~\cite{asatryan2013wide} \\
   $\alpha$ (deg), site 1
     & -35.08(10) & 61.52(11) & -35~\cite{asatryan2013wide} \\
   $\alpha$ (deg), site 2
     & 35.08(10) & -61.52(11) & 35~\cite{asatryan2013wide} \\
   $g_n \mu_N / h$ ($\mathrm{MHz/T}$)
     & -3.53 & -3.53 & -3.53~\cite{PhysRevB.104.134103} \\
 \end{tabular}
 \end{ruledtabular}
\end{table}

\section{\label{sec:optical_clock_transitions}Optical clock transitions}

\begin{table}[t]
\caption{Optical clock points for the ZPL transitions for site 1. For site 2, angle $\phi \rightarrow -\phi$. For each point it is verified that $S1\Delta B \ll S2(\Delta B)^2$ with $\Delta B=1~ \mathrm{mT}$.}
  \label{tab:optical_clocks_list}
  \begin{ruledtabular}
  \begin{tabular}{ccccc} %
    Transition  & $|\mathbf{B}|$ (mT) & $\theta$ (deg) & $\phi$ (deg) & S2 (Hz/G$^2$) \\

    \colrule
    \multirow{7}{*}{Z1 $\uparrow$$\leftrightarrow$Y1 $\uparrow$}
& $12.29$ & $0.0$ & $\diagup$ & $1.7\cdot 10^{6}$\\
& $12.35$ & $90.0$ & $145.0$ & $1.6\cdot 10^{6}$\\
& $14.58$ & $90.0$ & $61.5$ & $1.3\cdot 10^{3}$\\
& $18.57$ & $90.0$ & $16.4$ & $1.3\cdot 10^{3}$\\
& $19.92$ & $47.2$ & $61.8$ & $1.3\cdot 10^{3}$\\
& $19.92$ & $132.8$ & $61.8$ & $1.3\cdot 10^{3}$\\
& $20.56$ & $90.0$ & $100.1$ & $1.3\cdot 10^{3}$\\

    \colrule
    \multirow{1}{*}{Z1 $\uparrow$$\leftrightarrow$Y1 $\downarrow$}
& $14.64$ & $90.0$ & $61.3$ & $1.3\cdot 10^{3}$\\

    \colrule
    \multirow{2}{*}{Z1 $\downarrow$$\leftrightarrow$Y1 $\uparrow$}
& $14.24$ & $90.0$ & $145.0$ & $1.4\cdot 10^{6}$\\
& $16.75$ & $0.0$ & $\diagup$ & $1.2\cdot 10^{6}$\\

    \colrule
    \multirow{3}{*}{Z1 $\uparrow$$\leftrightarrow$X1 $\uparrow$}
& $14.43$ & $90.0$ & $54.2$ & $1.3\cdot 10^{3}$\\
& $20.79$ & $0.0$ & $\diagup$ & $9.8\cdot 10^{5}$\\
& $20.79$ & $90.0$ & $145.0$ & $9.8\cdot 10^{5}$\\

    \colrule
    \multirow{1}{*}{Z1 $\uparrow$$\leftrightarrow$X1 $\downarrow$}
& $14.82$ & $90.0$ & $56.2$ & $1.3\cdot 10^{3}$\\

    \colrule
    \multirow{2}{*}{Z1 $\downarrow$$\leftrightarrow$X1 $\uparrow$}
& $27.84$ & $0.0$ & $\diagup$ & $7.3\cdot 10^{5}$\\
& $49.37$ & $90.0$ & $145.0$ & $4.1\cdot 10^{5}$\\

    \colrule
    \multirow{6}{*}{Y1 $\uparrow$$\leftrightarrow$X1 $\uparrow$}
& $11.75$ & $90.0$ & $145.1$ & $1.2\cdot 10^{2}$\\
& $22.67$ & $90.0$ & $61.4$ & $49$\\
& $27.58$ & $90.0$ & $32.4$ & $16$\\
& $30.19$ & $90.0$ & $87.2$ & $16$\\
& $30.88$ & $49.0$ & $61.8$ & $8.9$\\
& $30.88$ & $131.0$ & $61.8$ & $8.9$\\

    \colrule
    \multirow{2}{*}{Y1 $\uparrow$$\leftrightarrow$X1 $\downarrow$}
& $16.71$ & $90.0$ & $145.1$ & $97$\\
& $81.29$ & $0.0$ & $\diagup$ & $16$\\

    \colrule
    \multirow{1}{*}{Y1 $\downarrow$$\leftrightarrow$X1 $\uparrow$}
& $31.57$ & $90.0$ & $61.3$ & $66$\\

    \colrule
    \multirow{1}{*}{Y1 $\downarrow$$\leftrightarrow$X1 $\downarrow$}
& $2.31$ & $0.0$ & $\diagup$ & $2.5\cdot 10^{2}$\\

  \end{tabular}
  \end{ruledtabular}
\end{table}

As discussed in Sec.~\ref{sec:introduction}, extending the optical coherence time of the X1-Y1 ZPL in Tm\ts{3+}:YAlO\bs{3} would be particularly exciting because this may allow one to develop quantum memory and indistinguishable single-photon sources at telecommunication wavelength. In a previous study, we found that the measured coherence times were mainly limited by local magnetic field fluctuations caused by spin flip-flops \cite{li2026spectroscopy}. The fully characterized spin Hamiltonians of the X1 and Y1 levels now enable us now to predict how each level splits and shifts under an arbitrary applied magnetic field, and hence to predict magnetic fields for which the frequency of a transition $\nu_t$ is unaffected by small changes of both the field's amplitude and direction. These particular fields are referred-to as optical clock points \cite{PhysRevA.85.032339}.

\begin{figure}[t]
    \centering
    \includegraphics[width=0.9\linewidth]{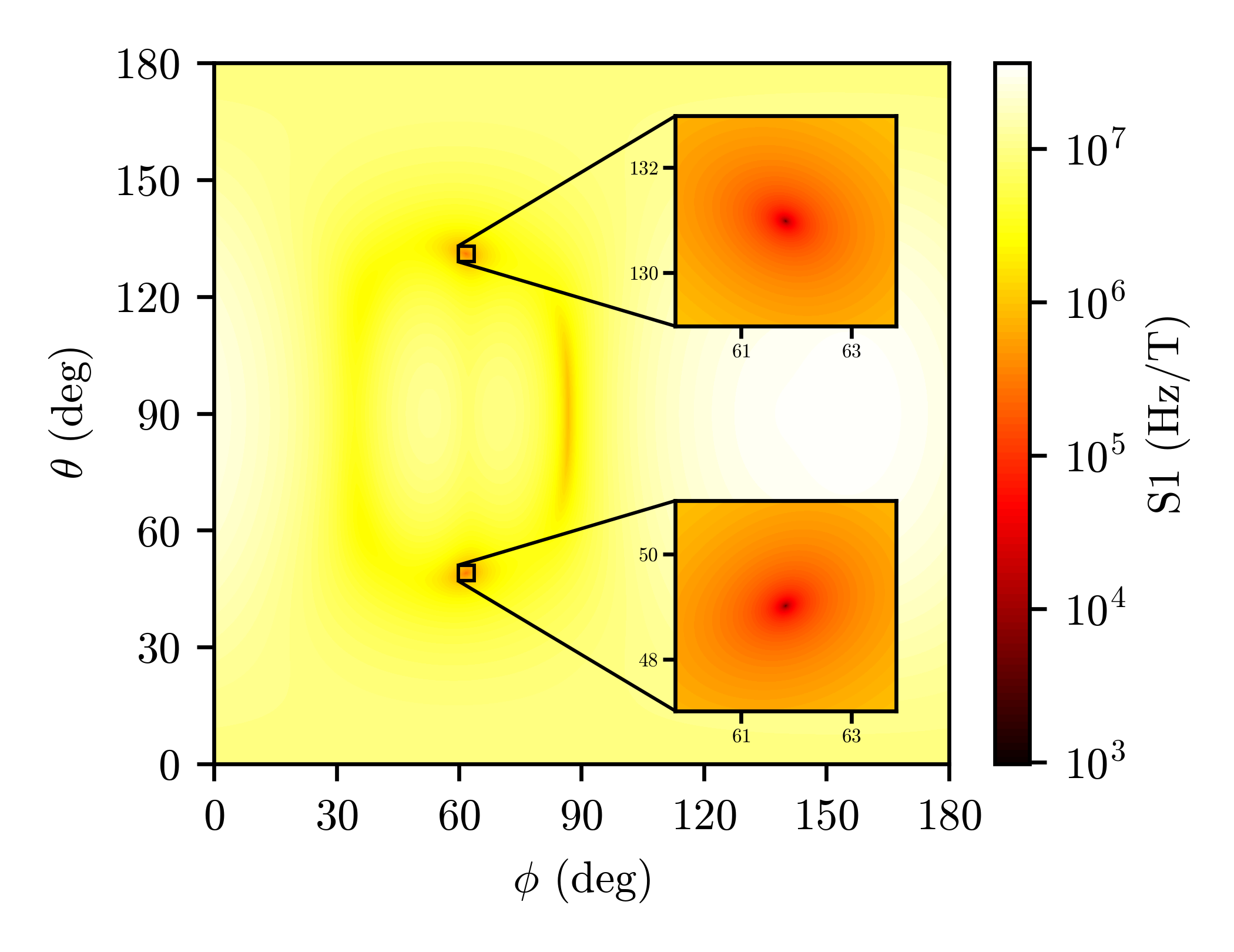}
    \caption{Heatmap of the first-order magnetic sensitivity $S_1$ for the Y1$\uparrow$ $\leftrightarrow$ X1$\uparrow$ transition at field amplitude $B=30.88$ mT. The $(\theta, \phi) = (49.0^\circ, 61.8^\circ)$ and $(\theta, \phi)=(131.0^\circ, 61.8^\circ)$ points both exhibit exceptionally low first order sensitivities below $10^{-2}$ Hz/T, indicating the presence of optical clock transitions. }
    \label{fig:S1_map}
\end{figure}

To determine these optical clock points systematically, we define the first-order sensitivity to magnetic noise S1 as:
\begin{equation}
    S1=|\nabla_{\mathbf{B}}\nu_t|.
\end{equation}
\noindent
S1 was computed numerically for the optical transitions between X1, Y1 and Z1, on angular grids of 1$^\circ$ resolution for all magnetic field orientations, using the Hamiltonian parameters from Table \ref{tab:spin_hamiltonian_params_global_fit}. This resulted in heatmaps such as in Fig.~\ref{fig:S1_map}, which we calculated for magnetic field amplitudes ranging from 0 to 10 T in 1 mT steps. Local minima of S1 were extracted, and we checked whether these represent true optical clock points with a root-finding algorithm, following \cite{PhysRevB.74.195101}. The optical clock points are the magnetic fields for which all three partial derivatives $\partial_B\nu_t$, $\partial_\theta\nu_t$ and $\partial_\phi\nu_t$ vanish simultaneously; more details are available in Appendix \ref{app:optical clocks}. The optical clock points found for Tm\ts{3+}:YAlO\bs{3} using this technique are listed in Table \ref{tab:optical_clocks_list}.

At an optical clock point, the second-order magnetic sensitivity becomes the dominant decoherence term. Under this assumption%
, the coherence time $T_2$ can be roughly estimated following a simple dephasing model \cite{PhysRevB.74.195101}:
\begin{equation}
\label{eq:coherence_optical_clock}
    \frac{1}{T_2}\approx S_2(\Delta B)^2,
\end{equation}
where $\Delta B$ is the standard deviation of the magnetic fluctuations, and $S2$ is the second order magnetic sensitivity. The latter is equal to the largest of the absolute values of the eigenvalues of the Hessian of $\nu_t$, and characterizes the curvature of $\nu_t$. See Appendix \ref{app:optical clocks} for additional information. 

As shown in Table \ref{fig:S1_map}, the value of S2 varies substantially between different optical clock points. %
As an example, the S2 value of the Y1$\uparrow$ $\leftrightarrow$ X1$\uparrow$ transition, addressed at the ($|B|=30.88\text{ mT, }\theta=131.0^\circ\text{, } \phi=61.8^\circ$) clock point, is 13 times smaller than that at the ($|B|=11.75\text{ mT, }\theta=90^\circ$\text{, }$\phi=145.1^\circ$) point. Consequently, according to Eq.~\eqref{eq:coherence_optical_clock}, the coherence time should be more than 13 times larger. Note that all optical clock transitions involving the Z1 level of the \Level{3}{H}{6} manifold have much larger S2, making them less suitable for coherence applications.

\section{\label{sec:conclusion}Conclusions}
We have characterized the spin Hamiltonians of the lowest Stark levels Y1 and X1 of the \Level{3}{F}{4}) and \Level{3}{H}{4}) manifolds of Tm\ts{3+}:YAlO\bs{3}, which are connected by a zero-phonon line at $\sim$1451 nm. Both levels correspond to optically excited states, and the characterization was carried out entirely by optical means and can be generalized to other crystals. The enhanced nuclear Zeeman splitting was obtained from the detuning of side holes created by spectral hole burning, and the quadratic Zeeman shift was calculated using the field-dependent displacement of the inhomogeneous line. The dependence of both measurements on the orientation of the magnetic field was well fitted by the spin-Hamiltonian model, yielding the $\mathbf{\Lambda}_l$ tensors for both levels and sites as well as the hyperfine constants $A_l$ for both manifolds. 

Using the spin Hamiltonians, together with literature values for the Z1 level of the \Level{3}{H}{6} ground-state manifold, we predict optical clock points at which the transition frequency is first-order insensitive to magnetic-field fluctuations for the Z1 $\leftrightarrow$ X1, Z1 $\leftrightarrow$ Y1 and Y1 $\leftrightarrow$ X1 transitions. The clock points of the excited-state Y1 $\leftrightarrow$ X1 transition are of particular interest. Indeed, since the optical coherence of the 1451 nm transition is currently limited by magnetic field fluctuations arising from spin flip-flops, operating at such a point is expected to extend it. In turn, this would open the path to long-lived quantum memory and indistinguishable single-photon sources at telecommunication wavelength.

\begin{acknowledgments}
We thank Patrick Remy, Stefan Stutz, and Maël Clémence for experimental help and discussions. WT acknowledges funding through the Swiss State Secretariat for Education, Research and Innovation (SERI) under contract No UeM019-3 and the Swiss National Science Foundation (project number 200021-236657).TS and GA acknowledge funding from the European Research Council under the European Union’s Horizon 2020 research and innovation program, within Grant Agreement 810451 (HERO).
\end{acknowledgments}

\section*{Data availability}
The data that support the findings of this article are openly available \cite{karyath_2026_21994306}.

\appendix
\setcounter{figure}{0}
\renewcommand{\thefigure}{A\arabic{figure}}
\renewcommand{\theHfigure}{appendix.A\arabic{figure}}
\section{Perturbation theory on spin Hamiltonian}
\label{app:perturbation theory}
We consider the lowest Stark level $\ket{0}$ in any manifold with unperturbed energy $E_0$, satisfying

\begin{equation}
    \left(H_{\mathrm{FI}}+H_{\mathrm{CF}}\right)\ket{0}
    =
    E_0\ket{0},
\end{equation}
where the index FI labels the free-ion Hamiltonian and CF the crystal field Hamiltonian.
The other Stark levels are denoted by $\ket{n}$, with unperturbed energies $E_n$. The following derivation is performed separately for each Stark level of interest, such as X1 or Y1. 

The first-order effective Hamiltonian associated with $\ket{0}$ is

\begin{equation}
    H_{\mathrm{eff}}^{(1)}
    =
    \bra{0}H_{\mathrm{pert}}\ket{0},
\end{equation}
where $H_{\mathrm{pert}}$ is defined in Eq.~\eqref{eq:pert_Hamiltonian} in the main text. Using this expression leads to

\begin{equation}
\begin{aligned}
    H_{\mathrm{eff}}^{(1)}
    ={}&
    A\mathbf{I}\cdot\bra{0}\mathbf{J}\ket{0}
    +
    g_J\mu_B\bra{0}\mathbf{J}\ket{0}\cdot\mathbf{B}
    \\
    &-
    g_n\mu_N\mathbf{I}\cdot\mathbf{B}.
\end{aligned}
\end{equation}

Because thulium is a non-Kramers ion, each Stark level is a non-degenerate singlet. Time-reversal symmetry
requires

\begin{equation}
    \bra{0}\mathbf{J}\ket{0}=0.
\end{equation}
\noindent
Consequently, the hyperfine and electronic Zeeman interactions vanish at first order, and only the nuclear Zeeman interaction remains:

\begin{equation}
    H_{\mathrm{eff}}^{(1)}
    =
    -g_n\mu_N\mathbf{I}\cdot\mathbf{B}.
\label{eq:appendix_first_order}
\end{equation}

We now consider the second-order contribution of the hyperfine and electronic Zeeman interactions and define
\begin{equation}
    V
    =
    A\mathbf{I}\cdot\mathbf{J}
    +
    g_J\mu_B\mathbf{J}\cdot\mathbf{B}.
\label{eq:appendix_V}
\end{equation}
The second-order contribution to the effective Hamiltonian is then
\begin{equation}
    H_{\mathrm{eff}}^{(2)}
    =
    -\sum_{n\neq 0}
    \frac{
        \bra{0}V\ket{n}
        \bra{n}V\ket{0}
    }{
        E_n-E_0
    }.
\label{eq:appendix_second_order}
\end{equation}
Plugging  Eq.~\eqref{eq:appendix_V} into Eq.~\eqref{eq:appendix_second_order}, we find
\begin{equation}
\begin{aligned}
H_{\mathrm{eff}}^{(2)}
={}&
-\sum_{n\neq 0}
\sum_{\alpha,\beta}
\frac{
    \bra{0}J_\alpha\ket{n}
    \bra{n}J_\beta\ket{0}
}{
    E_n-E_0
}
\\
&\times
\Bigl[
    g_J^2\mu_B^2 B_\alpha B_\beta
    +
    A^2 I_\alpha I_\beta
\\
&\qquad
    +
    A g_J\mu_B
    \left(
        B_\alpha I_\beta
        +
        I_\alpha B_\beta
    \right)
\Bigr],
\end{aligned}
\label{eq:appendix_second_order_components}
\end{equation}
where $\alpha, \beta \in \{x,y,z\}$. Using the $\Lambda$ tensor defined in Eq.~\eqref{eq:lambda_tensor} in the main text, we find

\begin{equation}
\begin{aligned}
H_{\mathrm{eff}}^{(2)}
={}&
-\sum_{\alpha,\beta}
\Lambda_{\alpha\beta}
\Bigl[
    g_J^2\mu_B^2 B_\alpha B_\beta
    +
    A^2 I_\alpha I_\beta
\\
&\qquad\qquad
    +
    A g_J\mu_B
    \left(
        B_\alpha I_\beta
        +
        I_\alpha B_\beta
    \right)
\Bigr].
\end{aligned}
\label{eq:appendix_second_order_lambda}
\end{equation}

Because $\boldsymbol{\Lambda}$ is real and symmetric, Eq.~\eqref{eq:appendix_second_order_lambda} can be written in matrix form as
\begin{equation}
\begin{aligned}
H_{\mathrm{eff}}^{(2)}
={}&
-g_J^2\mu_B^2
\mathbf{B}^{\mathsf T}\boldsymbol{\Lambda}\mathbf{B}
\\
&-
2A g_J\mu_B
\mathbf{B}^{\mathsf T}\boldsymbol{\Lambda}\mathbf{I}
\\
&-
A^2
\mathbf{I}^{\mathsf T}\boldsymbol{\Lambda}\mathbf{I}.
\end{aligned}
\label{eq:appendix_second_order_final}
\end{equation}

For $I=1/2$, the last term is proportional to the identity:
\begin{equation}
\mathbf{I}^{\mathsf T}\mathbf{\Lambda}\mathbf{I}
=
\frac{1}{4}\operatorname{Tr}(\mathbf{\Lambda})
\mathds{1}_{I}.
\end{equation}
It therefore produces only a field-independent common shift of the nuclear-spin sublevels. We absorb this contribution into the zero-field energy and omit it from the field-dependent effective Hamiltonian.

Therefore, by combining the first-order and second-order contributions of the perturbation Hamiltonian, we obtain the effective spin Hamiltonian in Eq.~\eqref{eq:perturbation_theory_Hamiltonian} in the main text.

\section{Leveraging symmetry to constrain tensor arithmetic}
\label{app:sym details}

\begin{figure}[t]
    \centering
    \includegraphics[width=0.9\linewidth]{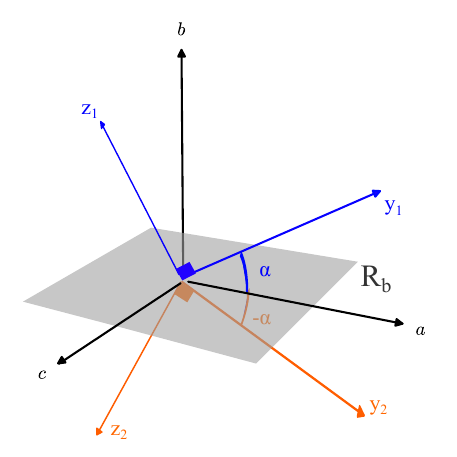}
    \caption{The mirror symmetry between site 1 and site 2 connects their diagonalization bases. The mirror $\hat{R_b}$ is shown in grey.}
    \label{fig:siteconnection}
\end{figure}

\begin{figure*}[t]
    \centering
    \includegraphics[width=1.3\columnwidth]{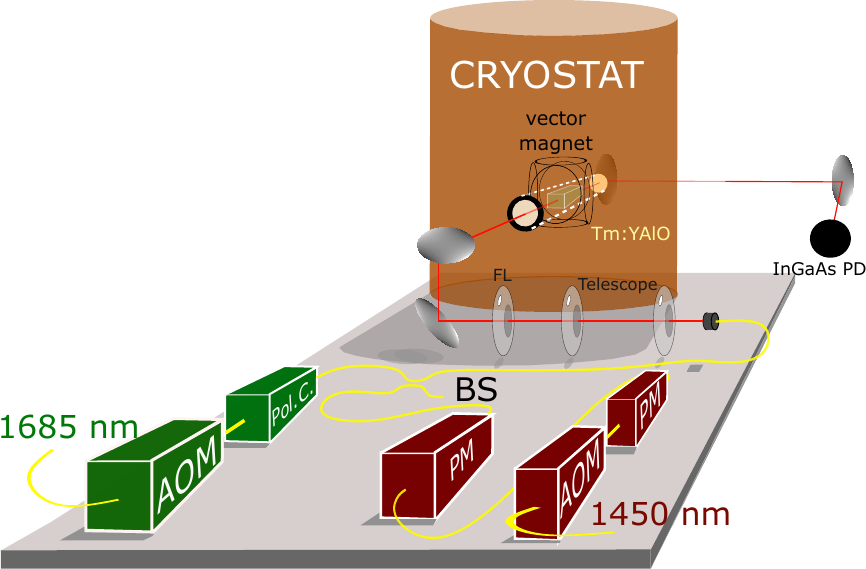}
    \caption{Experimental set-up. Fibered polarization controller (Pol.C) allows to tune the polarization of both pump and probe beams. Acousto-optic modulators (AOM) and phase modulators (PM) allow to shape the light in time and frequency. Both pump and probe beam are recombined using a fibered beam splitter (BS). The periscope, the telescope and the lens (FL) on translation stages offer the degrees of freedom required to focus the light in the crystal through the cryostat, and collect it. }
    \label{fig:setup}
\end{figure*}

This part provides mathematical details for the predictions presented in Section \ref{sec:crystal_symmetry}.

In our case, the Neumann principle states that any physical property of an ion must exhibit at least the symmetry of the point group of its site. Mathematically, if a site is invariant under some symmetry operation $\hat{S}$, any polar vector $V$ representing an intrinsic physical property at this site must satisfy:
\begin{align}
    V=\hat{S}V
    \label{eq:symvec}
\end{align}
For a second-rank polar tensor, such as the $\mathbf{\Lambda}$ tensor, the Neumann principle takes the form:
\begin{align}
    \Lambda=\hat{S}\Lambda \hat{S}^\top
    \label{eq:symten}
\end{align}
In YAlO$_3$ crystals where rare-earth ions occupy sites with $C_s$ point symmetry (a single mirror plane orthogonal to the $c$ axis), the symmetry operator $S$ is expressed as:
\begin{align}
    \hat{S}=\begin{pmatrix}
        -1 & 0 & 0\\
        0 & 1& 0\\
        0 & 0 & 1
    \end{pmatrix}_{(c,a,b)}
    \label{eq:symyal}
\end{align}
The tensor presented in Eq. \ref{eq:one}, in the $(c,a,b)$ crystal reference frame, is obtained by combining Eq. (\ref{eq:symten}) and Eq. (\ref{eq:symyal}). The mirror plane symmetry enforces $c$ to be a principal axis of any second-rank polar tensor describing the physical properties of ions located at this site of YAlO$_3$.

In addition, it is obvious from Eq. (\ref{eq:lambda_tensor}) in the main text that $\Lambda_{l_{a,b}}=\Lambda_{l_{b,a}}$. Because the $\mathbf{\Lambda}$ tensor is real and symmetric, it can be diagonalized in an orthonormal basis (x,y,z), as expressed in Eq. (\ref{eq:one}).

The other simplification discussed in the paper is given by the glide plane mirror connecting sites 1 and 2. This symmetry is depicted in Fig.~\ref{fig:siteconnection}, it is the combination of a mirror reflection $\hat{R_b}$ and a spatial translation $\hat{T}$.

Because directional physical quantities and tensors transform strictly under directional changes rather than spatial translations, the translational component $\hat{T}$ acts as the identity on tensor components. Consequently, only the reflection operator $\hat{R}$ dictates the tensor transformation rules between the two sites:
\begin{align}
    \hat{R_b}=\begin{pmatrix}
        1 & 0 & 0\\
        0 & 1& 0\\
        0 & 0 & -1
    \end{pmatrix}_{(c,a,b)}
    \label{eq:symmir}
\end{align}

\begin{align}
    \mathbf{\Lambda}^1=\hat{R_b} \mathbf{\Lambda}^2 \hat{R_b}^\top
    \label{eq:symsite}
\end{align}
To understand the geometric orientation of the local axes, let $\mathbf{y}_1$ be the unit vector along the local $y$ axis of site 1, forming an angle $\alpha$ with the crystal $a$ axis in the $ab$ plane:
\begin{align}
    \mathbf{y}_1 = \begin{pmatrix} 0 \\ \cos\alpha \\ \sin\alpha \end{pmatrix}_{(c,a,b)}
\end{align}
Applying the reflection operator $\hat{R_b}$ yields the local $y$-axis direction for site 2, $\mathbf{y}_2$:
\begin{align}
    \mathbf{y}_2 = \hat{R_b} \mathbf{y}_1 = \begin{pmatrix} 0 \\ \cos(-\alpha) \\ \sin(-\alpha) \end{pmatrix}_{(c,a,b)}
\end{align}
This explicitly demonstrates that the local $y$ axis of site 2 forms an angle of $-\alpha$ with the crystal $a$ axis. Note that the transformation reverses the orientation of the local coordinate system, mapping a right-handed basis at site 1 onto a left-handed basis at site 2. However, since $\Lambda$ is a second-rank polar tensor, its transformation properties are invariant under parity inversion, and this has no physical effect on the tensor components or its principal values.

\section{Experimental setup}
\label{app:set up}
A schematic of the experimental setup described in Section IV is depicted in Fig. \ref{fig:setup}. 

\section{Calibration of magnetic field}

\label{app:calib}
For our studies, it is crucial to know the direction of the magnetic field with respect to the crystal axes ($a$, $b$, $c$). However, due to an unavoidable misalignment of the crystal inside the cryostat, the reference system used by the software to create a desired magnetic field and the crystal axes differ slightly. Hence, an important task is to characterize the transformation from one to the other reference frame. Towards this end, we exploited that the particular symmetry of the crystal constrains the 2 Tm\ts{3+}sites to become magnetically equivalent for magnetic fields within the $ac$ and $bc$ planes, see Fig. \ref{fig:enhanced_nuclear_Zeeman}.

\begin{figure}[htbp]
    \centering
    \includegraphics[width=0.9\linewidth]{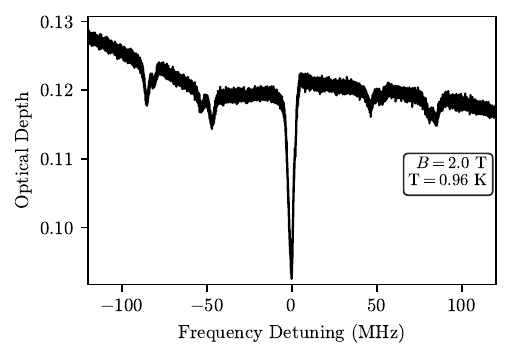}
    \caption{With the external magnetic field imperfectly aligned with the $b$ axis, the magnetically inequivalent sites exhibit slightly different detunings of the side holes.}
    \label{fig:merging}
\end{figure}

We used the following method: 1) the magnetic field was oriented roughly along the $b$ axis. At this point, the two sites could be observed but their spectra almost overlapped (see Fig.~\ref{fig:merging}); 2) the angle $\phi$ (see Fig.~\ref{fig:coordinate_system}) of the magnetic field was tuned until the spectral holes coincided, indicating that B was now oriented within the $bc$ plane (see Fig.~\ref{fig:sideholes} for an example); 3) the angle $\theta$ (see Fig.~\ref{fig:coordinate_system}) of the magnetic field was changed by approximately 30 degrees and the merging procedure was performed a second time ; 4) at this point we know 2 vectors in the magnet reference frame that lie within the $bc$ plane of the crystal. Taking the vector product of these returns the $a$ axis ; 5) The same procedure is repeated in the $ac$ plane, allowing us to deduce the $b$ axis  ; 6) the vector product of $a$ and $b$ yields $c$, which allows us to compute a transformation matrix to pass from the magnet reference frame to the crystal reference frame and to label all magnetic field directions using the latter.

\section{Experimental identification of two magnetically inequivalent sites\label{app:site_convention}}
As discussed in Sec.~\ref{sec:SHB}, the two sites occupied by thulium ions are magnetically inequivalent unless the applied magnetic field is oriented within the $ac$ or $bc$ planes. Therefore, we need to establish a convention to assign each measured splitting to a specific site when they are magnetically inequivalent.

We start by applying the magnetic field along the $b$ axis of the crystal, in which case we observe that the two sites are magnetically equivalent and that their side holes coincide. Then, we rotate the magnetic field within the $ab$ plane in small steps away from the $b$ axis. As shown in  Figs.~\ref{fig:enhanced_nuclear_Zeeman} (a) and (d) in the main text, there are more data points around $\phi=90^{\circ}$. When the applied magnetic field is slighted misaligned with the $b$ axis, we can observe in total eight side holes (four side holes for each site). Because of the slight misalignment along the $b$ axis, the difference of detunings between the side holes caused by the two sites are small, and all the holes can be observed within a single readout.

As we reduce the angle $\phi$ from $90^{\circ}$ to $85^{\circ}$, one group of side holes show an increasing $\delta e_{\mathrm{X1}}$ and decreasing $\delta e_{\mathrm{Y1}}$; we label this group as site 1. The other group behaves oppositely and is labeled as site 2.

Following this convention, when $\mathbf{B}$ is around the $b$ axis, we can trace the detuning of side holes and assign them to site 1 or 2. We use Eq.~\eqref{eq:splitting_value} to fit these measurements. By applying the fitted parameters back to Eq.~\eqref{eq:splitting_value}, we can predict the enhanced nuclear Zeeman splitting when $\mathbf{B}$ is applied along other orientations within the $ab$ plane. Based on the predictions, we can attribute the other measurements when $\mathbf{B}$ is within $ab$ plane to site 1 or site 2.

We also point out that there exist orientations for which the quadratic Zeeman shift rates differ significantly between site 1 and site 2, see Fig.~\ref{fig:orient shift} (a). For example, when $\mathbf{B}$ is applied along $\theta=90^{\circ},~\phi=30^{\circ}$, the shift rates $\Delta k$ are around -131 and +861 $\mathrm{MHz/T^2}$ for site 1 and site 2, respectively. When the magnitude of the applied field exceeds 1.1 T, the $\sim$1.2 GHz-wide inhomogeneous broadened line splits completely, and the ions can be probed separately. Similarly, when considering the Z1-Y7 transition used for optical pumping, it is also possible to address the two sites separately.

\section{Quality of the fits}
\label{app:fit}
The residual standard error (RSE) is defined as
\begin{equation}
    \mathrm{RSE}=\sqrt{\sum_{i}(y_i-f_i)^2/(N-n_{\text{par}})},
\end{equation}
where $y_i$ is the measured value, $f_i$ is the fitted value, $N$ is the number of data points, and $n_{\text{par}}$ is the number of fitted parameters. 

For the enhanced nuclear Zeeman splitting, the fits shown in Fig.~\ref{fig:enhanced_nuclear_Zeeman} result in an RSE of 0.342 MHz/T for the X1 level and of 0.408 MHz/T for the Y1 level. Both are  small compared to the scale of the measured splitting rates ($9.5 - 50.3~\mathrm{MHz/T}$), indicating that the angular dependence of the splitting rates of both X1 and Y1 is well described by Eq.~\eqref{eq:splitting_value}.

For the quadratic Zeeman shift, the RSE of the fit shown in Fig.~\ref{fig:orient shift} in the main text is $10.6~\mathrm{MHz/T^2}$, again small compared to the measured shift rates whose absolute values all exceed 100 $~\mathrm{MHz/T^2}$. As above, this shows 
that the measurement results are well described by Eq.~\eqref{eq:shift value}.

\section{Optical clock transitions}
\label{app:optical clocks}
\begin{figure*}[htbp]
    \centering
    \includegraphics[width=0.8\textwidth]{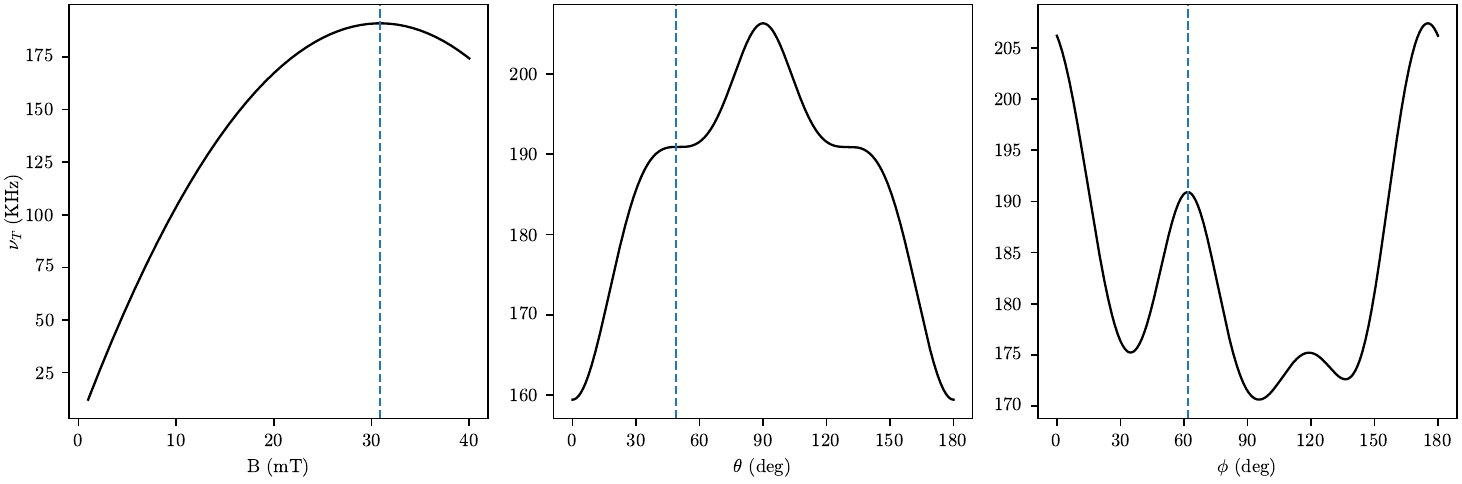}
    \caption{Computed magnetic-field dependent frequency detuning $\nu_t$ of the Y1 $\uparrow$ $\leftrightarrow$ X1 $\uparrow$ optical transition for variations around the $(B, \theta, \phi)=(30.88\text{ mT},49.0^\circ,61.8^\circ)$ optical clock point. The vertical blue dashed lines indicated the setting where the optical clock occurs, and the slope is zero.}
    \label{fig:optical_clock_example}
\end{figure*}

It is instructive to plot the energy as a function of $|B|$, $\theta$ and $\phi$ at an optical clock point. A typical example is given in  Figure \ref{fig:optical_clock_example} ; it is the visual representation of a solution to the equation $S1=|\nabla_{\textbf{B}} \nu_t|= 0$. To numerically locate clock points, a least square minimization of the gradient norm is programmed over the parameters $\phi,\theta,|B|$. To ensure numerical stability and rapid convergence, both the gradient and its exact Jacobian are evaluated via automatic differentiation rather than finite differences. This gradient-based formulation allows us to determine critical field orientations and amplitudes with machine-level precision (tolerances set to $10^{-14}$).

The parameter S2 is related to the curvature of the plots shown in Fig.~\ref{fig:optical_clock_example}; formally it is defined as the maximum of the eigenvalues of the Hessian matrix of $\nu_t$:
\begin{align}
    S2=\max(\lambda_i(H_{\nu_t}))
    \label{eq:s2}
\end{align}
where the Hessian matrix, expressed in spherical coordinates, is given by:
\begin{equation}
    H_{\nu_t} = 
    \begin{pmatrix}
        (H_{\nu_t})_{\phi\phi} & (H_{\nu_t})_{\phi\theta} & (H_{\nu_t})_{\phi |B|} \\
        (H_{\nu_t})_{\theta\phi} & (H_{\nu_t})_{\theta\theta} & (H_{\nu_t})_{\theta |B|} \\
        (H_{\nu_t})_{|B|\phi} & (H_{\nu_t})_{|B|\theta} & (H_{\nu_t})_{|B||B|}
    \end{pmatrix}.
    \label{eq:hessian}
\end{equation}

The Hessian is symmetric, $(H_{\nu_t})_{ij}=(H_{\nu_t})_{ji}$, with
independent components
\begin{subequations}
\label{eq:hessian-spherical}
\begin{align}
(H_{\nu_t})_{\phi\phi} &= \frac{\partial_{\phi}^{2}\nu_t}{B^{2}\sin^{2}\theta}
  + \frac{\cos\theta}{B^{2}\sin\theta}\,\partial_{\theta}\nu_t
  + \frac{1}{B}\,\partial_{B}\nu_t, \\
(H_{\nu_t})_{\phi\theta} &= \frac{\partial_{\phi}\partial_{\theta}\nu_t}{B^{2}\sin\theta}
  - \frac{\cos\theta}{B^{2}\sin^{2}\theta}\,\partial_{\phi}\nu_t, \\
(H_{\nu_t})_{\phi B} &= \frac{\partial_{\phi}\partial_{B}\nu_t}{B\sin\theta}
  - \frac{1}{B^{2}\sin\theta}\,\partial_{\phi}\nu_t, \\
(H_{\nu_t})_{\theta\theta} &= \frac{1}{B^{2}}\,\partial_{\theta}^{2}\nu_t
  + \frac{1}{B}\,\partial_{B}\nu_t, \\
(H_{\nu_t})_{\theta B} &= \frac{1}{B}\,\partial_{\theta}\partial_{B}\nu_t
  - \frac{1}{B^{2}}\,\partial_{\theta}\nu_t, \\
(H_{\nu_t})_{BB} &= \partial_{B}^{2}\nu_t,
\end{align}
\end{subequations}
where $\partial_x \equiv \partial/\partial x$ and $B \equiv |\mathbf{B}|$. The value of S2 is calculated for the parameters that satisfy S1=0. Note that, because S2 is the largest component of the Hessian matrix, and $(\Delta B)^2$ is taken as the variance of the absolute value of the magnetic fluctuations, Eq.~\eqref{eq:coherence_optical_clock} therefore corresponds to a lower bound estimation of $T_2$. A vectorial version of Eq.~\eqref{eq:coherence_optical_clock} would be a most accurate version of the above development, it would nonetheless require a very accurate knowledge of the magnetic noise, which is not the case for the moment.

\FloatBarrier
\bibliography{references}

\end{document}